\documentstyle[12pt,epsf,rotate]{article}
\newbox\rotbox
\begin{document}


\centerline{\large\bf NEW QCD SUM RULES FOR IN-MEDIUM
NUCLEONS\footnote{Invited talk presented at Baryons '95 by Derek
Leinweber.}}
\vspace{16pt}
\centerline{DEREK B. LEINWEBER\footnote{
E-mail:~derek@phys.washington.edu ~$\bullet$~ Telephone:
(206)~616--1447 ~$\bullet$~ Fax: (206)~685--0635 \hfill\break
\null\qquad\quad WWW:~http://www.phys.washington.edu/$\sim$derek/}}
\centerline{\it Department of Physics, Box 351560, University of
Washington, Seattle WA 98195}
\vspace{14pt}
\centerline{XUEMIN JIN\footnote{E-mail:~jin@alph02.triumf.ca
{}~$\bullet$~ Telephone: (604) 222--1047 ext.\ 6446 ~$\bullet$~ Fax:
(604) 222--1074 }}
\centerline{\it TRIUMF, 4004 Wesbrook Mall, Vancouver, BC,
V6T 2A3, Canada}
\vspace{14pt}
\centerline{R.
J. FURNSTAHL\footnote{E-mail:~furnstah@mps.ohio-state.edu ~$\bullet$~
Telephone: (614) 292--4830 ~$\bullet$~ Fax: (614) 292--7557 \hfill\break
\null\qquad\quad WWW:~http://www-physics.mps.ohio-state.edu/$\sim$furnstah/}}
\centerline{\it Department of Physics, The Ohio State University,
Columbus, Ohio 43210}
\vspace{6pt}
\begin{quote}\hspace*{0.5cm}
\small
New QCD sum rules for nucleons in nuclear matter are derived from a
mixed correlation function of spin-1/2 and spin-3/2 interpolating
fields.  These sum rules allow a determination of the scalar
self-energy of the nucleon independent of the poorly known four-quark
condensates.  An analysis of these new sum rules in concert with
previous nucleon sum rules from spin-1/2 interpolators indicates
consistency with the expectations of relativistic mean-field
phenomenology.  We find $M^* = 0.68 \pm 0.08$ GeV and $\Sigma_v = 0.31
\pm 0.06$ GeV at nuclear matter saturation density. \\
\end{quote}

\section{Introduction}

   Understanding the observed properties of hadrons and nuclei from
quantum chromodynamics (QCD) is a principal goal of nuclear theorists.
The QCD sum-rule approach \cite{shifman1} is a particularly useful
method for connecting the properties of QCD to observed nuclear
phenomena.  Recent progress in understanding the origin of the large
and canceling isoscalar Lorentz-scalar and -vector self-energies for
propagating nucleons in nuclear matter has been made via the analysis
of QCD sum rules generalized to finite nucleon density
\cite{cohenr}.  These large self-energies
are central to the success of relativistic nuclear phenomenology.

   However, the previous sum-rule predictions for the scalar
self-energy are sensitive to the density dependence of certain
``scalar-scalar'' dimension-six four-quark condensates.  This density
dependence is unknown \cite{cohenr}.  There are at
least two ways to clarify the situation.  One direction is to attempt
to better determine the density dependence of the four-quark
condensates via modeling \cite{celenza1}.  An alternative approach,
which is adopted here, is to derive new QCD sum rules for the scalar
self-energy that do not depend on the four-quark condensates.

   The new sum rules are obtained from a mixed correlator of
generalized spin--1/2 and spin--$3/2$ interpolating fields.  The
spin--$1/2$ states remain projected, and one generates additional sum
rules for the scalar self-energy that are independent of the
problematic four-quark condensates.  These new sum rules, along with
previous sum rules from spin--1/2 interpolators that are also
independent of four-quark condensates \cite{cohenr},
provide an interesting forum for examining the expectations of
phenomenological approaches.  In this work we present results from the
first analysis of the new sum rules; more detailed analyses will be
given in future publications \cite{furnstahl2}.

\section{New QCD Sum Rules}

   The finite-density QCD sum-rule approach focuses on a correlation
function of interpolating fields carrying the quantum numbers of the
hadron of interest
\begin{equation}
\Pi_{\mu\nu}^{12}(q)\equiv i\int d^4 x\, e^{iq\cdot x}
\langle\Psi_0|T \left[ \chi^1_\mu(x) \, \overline{\chi}^2_\nu(0)
\right ]| \Psi_0 \rangle \, .
\label{corr_def}
\end{equation}
Here, the ground state of nuclear matter $|\Psi_0\rangle$ is
characterized by the rest-frame nucleon density $\rho_N$ and by the
four-velocity $u^\mu$.  The nucleon interpolating fields are
\cite{leinweber1,leinweber2}
\begin{eqnarray}
\chi^1_\mu(x)&=& \gamma_\mu \gamma_5 \epsilon^{abc}\left\{
\left[u_a^T(x) \, C \, \gamma_5 \, d_b(x) \right] u_c
+
\beta \left[u_a^T(x)\, C\, d_b(x)\right]\gamma_5\, u_c(x)
\right\}\, ,
\label{field-1}
\\*[7.2pt]
\chi^2_\nu(x)&=&\epsilon^{abc}\left\{
\left[u_a^T(x)\, C\, \sigma_{\rho\lambda}\, d_b(x) \right]
\sigma^{\rho\lambda} \, \gamma_\nu \, u_c(x)
-\left[u_a^T(x) \, C\, \sigma_{\rho\lambda}\, u_b(x) \right]
\sigma^{\rho\lambda} \, \gamma_\nu \, d_c(x) \right\} \, ,
\label{field-2}
\end{eqnarray}
where $T$ denotes a transpose in Dirac space, $C$ is the charge
conjugation matrix, and $\beta$ is a parameter allowing for arbitrary
mixing of the two independent spin-1/2 interpolators.  For $\beta=-1$,
the correlator of $\chi^1_\mu$ and $\overline{\chi}^1_\nu$ gives the
sum rules discussed in Ref.~\cite{cohenr}.

   In the rest frame of the medium, the analytic properties of the
invariant functions can be studied through Lehman representations in
energy.  The quasi-nucleon excitations are characterized by the
discontinuities of the invariant functions across the real axis, which
are used to identify the on-shell self-energies.  A representation of
the correlation function can be obtained by introducing a simple
phenomenological Ansatz for these spectral densities.

  On the other hand, the correlation function can be evaluated at
large space-like momenta using an operator product expansion (OPE).
This expansion requires knowledge of QCD Lagrangian parameters and
finite-density quark and gluon matrix elements.  Finite-density QCD
sum rules, which relate the nucleon self-energies in the nuclear
medium to these QCD inputs, are obtained by equating the two different
representations using appropriately weighted integrals \cite{cohenr}.

   The correlator $\Pi_{\mu\nu}^{12}(q)$ contains nine distinct
structures at finite density.  Two of these sum rules are independent
of the problematic ``scalar-scalar'' four-quark condensates and are
dominated by the leading terms of the OPE.

   To analyze the sum rules, we follow the techniques introduced in
Ref.~\cite{leinweber1,leinweber2} for determining the valid Borel
region.  We limit the continuum model contributions to 50\% of the
phenomenology, and maintain the contributions of the highest
dimensional operators in the OPE to less than 10\% of the sum of OPE
terms.  This defines a region in Borel parameter space where a sum
rule should be valid.  Uncertainties are estimated via a new
Monte-Carlo error analysis introduced in Ref.~\cite{leinweber2}.

   As in previous works on finite-density sum rules, we use the linear
density approximation for estimating the in-medium condensates.
Central values and uncertainties of QCD vacuum parameters are taken
{}from Ref.~\cite{leinweber2}.  Finite-density parameters are in accord
with the estimates of Ref.~\cite{cohenr}.  Since the mixed condensate
$\langle g_c \overline{q}\sigma\cdot G q\rangle_{\rho_N}$ is chirally
odd, we assume the same density dependence as for the quark
condensate.

   The one firm conclusion from previous in-medium studies is that the
vector self-energy is positive and a few hundred MeV.  Hence, we begin
by fixing $\Sigma_v = 0.25$ GeV at saturation density and searching
for a region in which both sides of the QCD sum rules are valid.

   There are two sum rules obtained from the correlator of $\chi^1_\mu$
and $\chi^1_\nu$ that are independent of the problematic four-quark
condensates.  One of these satisfies the validity criteria.
\begin{eqnarray}
\lambda_1^2 \, \Sigma_v \, e^{-(E_q^2-{\bf q}^2)/M^2}
&=&{5 + 2 \beta + 5 \beta^2 \over 48 \pi^2} \, M^4\,  E_1 \,
\langle q^\dagger q \rangle_{\rho_N} \, L^{-4/9}
\nonumber\\*
& &\null
+{5 \left (5 + 2 \beta + 5 \beta^2 \right ) \over 72 \pi^2} \,
 \overline{E}_q \, M^2 \, E_0 \,
 \langle q^\dagger iD_0 q \rangle_{\rho_N} \, L^{-4/9}
\nonumber\\*
& &\null
-{7 + 10 \beta + 7 \beta^2 \over 192 \pi^2} \, M^2\,  E_0\,
\langle g_c q^\dagger\sigma\!\cdot\!{\cal G} q\rangle_{\rho_N} \,
L^{-4/9}
\nonumber\\*
& &\null
+{5 + 2 \beta + 5 \beta^2 \over 8 \pi^2} \, {\bf q}^2
\left ( \langle q^\dagger iD_0 iD_0 q\rangle_{\rho_N}+{1 \over 12}
   \langle g_c q^\dagger\sigma\!\cdot\!{\cal G}q\rangle_{\rho_N}
\right ) \, L^{-4/9}
\nonumber\\*
& &\null
+{5 + 2 \beta + 5 \beta^2 \over 12} \, \overline{E}_q \,
\langle q^\dagger q\rangle_{\rho_N}^2 \, L^{-4/9}\ .
\label{sum-2}
\end{eqnarray}
Here $\lambda_1$ denotes the coupling of $\chi^1_\mu$ to the
quasi-nucleon state.  We have also defined
\begin{equation}
M_N^* \equiv M_N+\Sigma_s\ ,\hspace*{1cm}
E_q\equiv \Sigma_v+\sqrt{{\bf q}^2+M_N^{\ast 2}}\ ,\hspace*{1cm}
\overline{E}_q \equiv \Sigma_v-\sqrt{{\bf q}^2+M_N^{\ast 2}}\ ,
\end{equation}
where $\Sigma_s$ and $\Sigma_v$ are the scalar and vector
self-energies of the nucleon in nuclear matter, respectively.  The
anomalous dimensions of various operators have been taken into account
through the factor $L \equiv \ln(M^2/ \Lambda_{\rm
QCD}^2)/\ln(\mu^2/\Lambda_{\rm QCD}^2)$ \cite{shifman1}.  We have also
defined $E_0\equiv 1-e^{-s_0/M^2}$ and $E_1\equiv
1-e^{-s_0/M^2}\left(s_0/ M^2+1\right)$, which account for
excited-state contributions \cite{cohenr}.

Of the two favorable new sum rules obtained from the overlap of
spin-1/2 and spin-3/2 interpolators, only the sum rule at the
structure $\gamma_\mu \rlap{/}q q_\nu$, satisfies the validity
criteria.
\begin{eqnarray}
\lambda_1\, \lambda_2\, {1\over M_N^*}\, e^{-(E_q^2-{\bf q}^2)/M^2}
&=&
-{1 \over 8\pi^2} \, \langle\overline{q}q\rangle_{\rho_N}\,
M^2\, E_0\, L^{8/27}
-{3-\beta\over 64\pi^2}\,
\langle g_c \overline{q}\sigma\cdot G q\rangle_{\rho_N}\, L^{-2/27}
\nonumber
\\*[7.2pt]
&&+{3+5\beta\over 96}\, \langle\overline{q}q\rangle_{\rho_N}\,
\langle{\alpha_s\over \pi}\, G^2\rangle_{\rho_N}\, {1\over M^2}\,
L^{8/27} \, .
\label{sum-4}
\end{eqnarray}

   The parameter $\beta$ is selected to minimize continuum model
contributions while maintaining reasonable higher-dimension operator
contributions such that the pole may be resolved from the continuum
contributions \cite{leinweber3}.  The optimal choice for the sum rule
of (\ref{sum-2}) is $\beta = -0.7$.  The continuum model contribution
of (\ref{sum-4}) is independent of $\beta$.  Since the second term of
(\ref{field-1}) has little overlap with the ground state nucleon
\cite{leinweber3}, we set $\beta = 0$.

\newpage
\section{Implications}

   In vacuum, higher-dimensional condensates are typically evaluated
by adopting a factorization prescription.  It is generally accepted
that factorization fails for the four-quark operator.  However, the
dimension seven operator appearing in (\ref{sum-4}) appears to be
reasonably well approximated by the factorization prescription in
vacuum \cite{leinweber2}.  As a chirally odd operator its density
dependence should be qualitatively similar to that of the quark
condensate.  Fortunately, this behavior arises naturally in the
factorized form where the density dependence of the gluon condensate
is estimated to be a 7\% effect.  Hence the large uncertainties
associated with the density dependence of factorized operators are
largely eliminated in these new sum rules.  We will further examine
the sensitivity of these assumptions in a later work
\cite{furnstahl2}.

   The density dependence of the new sum rule of (\ref{sum-4}) is
predominantly governed by the quark condensate and is common to all
terms of the OPE.  As such, the effects of increasing density will be
to reduce the residue of the pole while the pole position remains
largely unchanged.  This result is the key feature absent in the
former in-medium nucleon analysis \cite{cohenr}.  The approximate
invariance of $M^* + \Sigma_v$ is manifest in (\ref{sum-4}).

   Fig. 1a displays the valid Borel regimes for the two sum
rules (\ref{sum-2}) and (\ref{sum-4}).  The distributions for $M^*$
and $\Sigma_v$ are illustrated in Fig. 1b.  We find $M^* =
0.68 \pm 0.08$ GeV and $\Sigma_v = 0.31 \pm 0.06$ GeV at nuclear
matter saturation density.  These results are consistent with the
expectations of relativistic phenomenology.

\begin{figure}[h]
\begin{center}
\epsfysize=6.4truein
\leavevmode
\setbox\rotbox=\vbox{\epsfbox{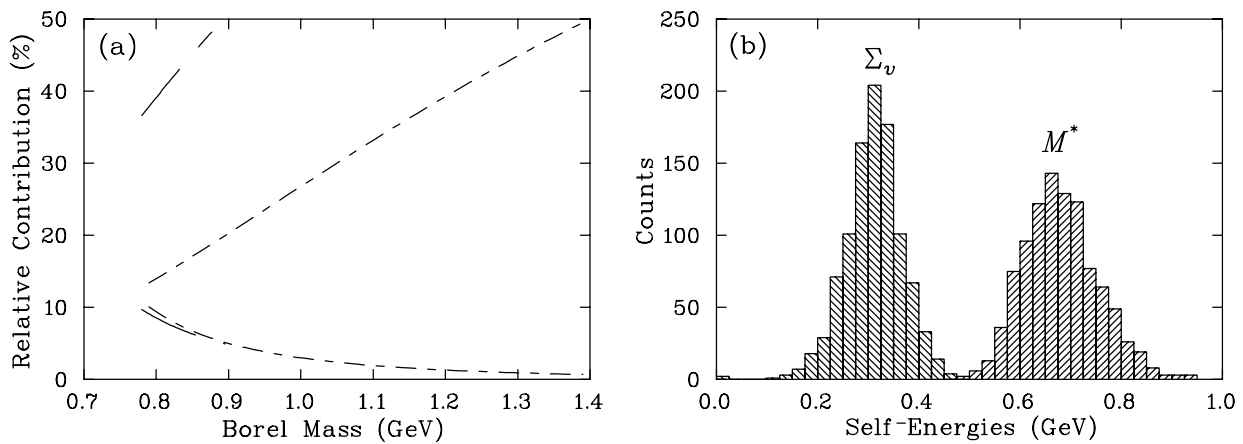}}\rotl\rotbox
\end{center}
\caption{\protect\small%
(a) Illustration of the valid Borel regimes for the sum rules
of (\protect\ref{sum-2}) (large dash) and (\protect\ref{sum-4}) (fine
dash).  Both continuum model contributions (limited to 50\%) and
highest dimension operator contributions (limited to 10\%) are
illustrated.  (b) Histogram for $\Sigma_v$ and $M^*$ obtained from a
Monte Carlo sample of 1000 QCD parameters.}
\label{fig-1}
\end{figure}


\end{document}